\documentclass[11pt]{article}
\usepackage{amsmath}
\usepackage[a4paper]{geometry}
\usepackage{graphicx}
\usepackage{color}
\usepackage{subfigure}
\usepackage[hidelinks]{hyperref}
\usepackage{url}
\usepackage{authblk}

\title{Signal model parameter scan using Normalizing Flow}

\author[1]{Masahiko Saito\thanks{E-mail: saito@icepp.s.u-tokyo.ac.jp}}
\author[1]{Masahiro Morinaga}
\author[2]{Tomoe Kishimoto}
\author[1]{Junichi Tanaka}

\affil[1]{International Center for Elementary Particle Physics (ICEPP), The University of Tokyo, 7-3-1, Hongo, Bunkyo, Tokyo, Japan}
\affil[2]{High Energy Accelerator Research Organization (KEK), Computing Research Center, 1-1, Oho, Tsukuba, Ibaraki, Japan}

\begin{document}
\bibliographystyle{unsrturl}
\maketitle

\begin{abstract}
The discovery of Beyond the Standard Model (BSM) is a major subject of many experiments, such as the ATLAS and CMS experiments with the Large Hadron Collider, which has the world's highest centre-of-mass energy.
Many types of BSM models have been proposed to solve the issues of the Standard Model.
Many of them have some or many model parameters, e.g. the Minimal Supersymmetric Standard Model, which is one of the most famous BSM models, has more than 100 model parameters.
These model parameters are free parameters; they cannot be predicted from theories and need to be determined experimentally.

Data analysis of BSM model searches involves comparing observed experimental data with a particular BSM model.
If the BSM model parameters are multidimensional, it is difficult to perform an analysis covering the whole phase space.
Instead, it is often performed by fixing all the model parameters except for one or two interesting parameters to focus on, or by using community defined benchmark points, resulting in phase space holes that are not covered by the search.

This paper presents a parameter scan technique for BSM signal models based on normalizing flow.
Normalizing flow is a type of deep learning model that transforms a simple probability distribution into a complex probability distribution as an invertible function.
By learning an invertible transformation between a complex multidimensional distribution, such as experimental data observed in collider experiments, and a multidimensional normal distribution, the normalizing flow model gains the ability to sample (or generate) pseudo experimental data from random numbers and to evaluate a log-likelihood value from multidimensional observed events.
The normalizing flow model can also be extended to take multidimensional conditional variables as arguments.
Thus, the normalizing flow model can be used as a generator and evaluator of pseudo experimental data conditioned by the BSM model parameters.
The log-likelihood value, the output of the normalizing flow model, is a function of the conditional variables.
Therefore, the model can quickly calculate gradients of the log-likelihood to the conditional variables.
Following this property, it is expected that the most likely set of conditional variables that reproduce the experimental data, i.e. the optimal set of parameters for the BSM model, can be efficiently searched.
This paper demonstrates this on a simple dataset and discusses its limitations and future extensions.
\end{abstract}

\section{Introduction}
The discovery of beyond the Standard Model (BSM) is one of the most important targets for high energy physics.
The LHC, as the highest energy collider, has been expected to discover BSM phenomena, but no evidence of BSM has been obtained so far, even after 15 years of operation.
Although many physics analyses have been performed~\cite{ATL-PHYS-PUB-2023-025, ATL-PHYS-PUB-2023-008,CMSSUSYSummaryPlots,CMSExoticaSummaryPlots}, not all possible BSM models and their phase space are covered.
To fully exploit the potential of the collected data, a more comprehensive physics analysis should be performed, not only focusing on a limited BSM phase space, such as benchmark points or simplified models.

One of the comprehensive physics analysis methods is one that uses anomaly detection.
In recent years, many methods have been proposed that use anomaly detection based on machine learning techniques~\cite{Kasieczka_2021, Aarrestad_2022}. Some of them search for anomalies using only observed data, i.e. unsupervised data, without assuming any signal model, or with assuming only a specific signal topology.
Such methods lead to improve the discovery sensitivity for BSM models that physicists did not anticipate, but worsen it for the known BSM models due to the lack of the assumptions on the models.

This study focuses on an analysis that assumes a specific signal model and explores the entire phase space of BSM model parameters, i.e., it is a signal model parameter scan.
The assumption of a specific BSM model results in a loss of sensitivity to other BSM models, but it is possible to explore an entire phase space that cannot be accessed by normal searches such as simplified models, and to search for the target BSM signals with more sensitivity than anomaly detection. 

Some BSM models have a very large number of model parameters.
For example, the Minimal Supersymmetric Standard Model (MSSM) has more than 100 parameters~\cite{DIMOPOULOS1981150}, and even its reduced model, the phenomenological MSSM, has 19 parameters~\cite{djouadi1999minimal}.
Exploring such a high-dimensional model parameter space with conventional methods is a difficult task.
The first reason is that as the dimension of the parameter space increases, the number of possible combinations increases exponentially. For example, grid search\footnote{The phase space is sliced on a grid and the values for each point on the grid are evaluated.} cannot be used when the dimension is larger than $\sim$3 because of the large number of combinations to be processed.
The curse of dimensionality can be avoided by using random search or Bayesian optimization.
However, since random search randomly samples points in phase space, it is difficult to estimate the best point unless the point close to the optimal point is accidentally sampled.
Bayesian optimization is able to efficiently sample points in phase space, but long waiting times occur for the signal parameter scan.
This is because the signal parameter scan requires a lot of computational resources and time for evaluation at each point in space; for the physics analysis case, it is necessary to generate Monte Carlo (MC) samples with specific model parameters and compare them with experimental data using appropriate evaluation metrics. In particular, MC sample generation takes several days, and tuning the evaluation strategy to maximize sensitivity requires time and effort.
Bayesian optimization is a sequential algorithm, so parallel evaluation is not possible, even though the total number of trials is small.
Therefore, a desired algorithm for such a parameter scan should be fast, handle high dimensions, and evaluate optimal values continuously.

We propose a method of signal parameter scan using Normalizing Flow (NF).
NF is a generative model and models a probability density function (pdf) from the unsupervised data.
In the parameter scan, the NF model is used as a generator of the BSM model by learning a distribution from the BSM model~($\{x_{\mathrm{obs}} \}$) with the BSM model parameters ($\theta_{\mathrm{BSM}}$) as conditional parameters, $x_{\mathrm{obs}^{'}} = f_{\mathrm{NF}}^{-1}(z|\theta_{\mathrm{BSM}})$ where $z$ is random variables, and is also used as an evaluator of the likelihood value based on the pdf of the observed data~(pdf $\propto \exp(-z^2/2)$, where $z = f_{\mathrm{NF}}(x_{\mathrm{obs}}|\theta_{\mathrm{BSM}})$).
Since NF is a neural network, the gradient of the likelihood value with respect to the BSM model parameters can be computed quickly by backpropagation.
By using the gradient values, the model parameter scan can be performed quickly even in high-dimensional spaces.
A similar idea has been proposed as simulation-based inference~\cite{Cranmer_2020}.
The novelty of the proposed method is to use gradients for fast parameter search in high-dimensional phase space.

An overview of the methodology is presented in Section~\ref{sec:workflow}, followed by application results based on toy data and LHC Olympic 2020 benchmark data in Section~\ref{sec:experiments}.

\section{Workflow} \label{sec:workflow}
The proposed method is performed in the following steps.

\begin{figure}[htbp]
  \begin{center}
    {
      \subfigure[Generates training samples]{%
        \includegraphics[width=0.33\columnwidth,bb=0 0 320 240]{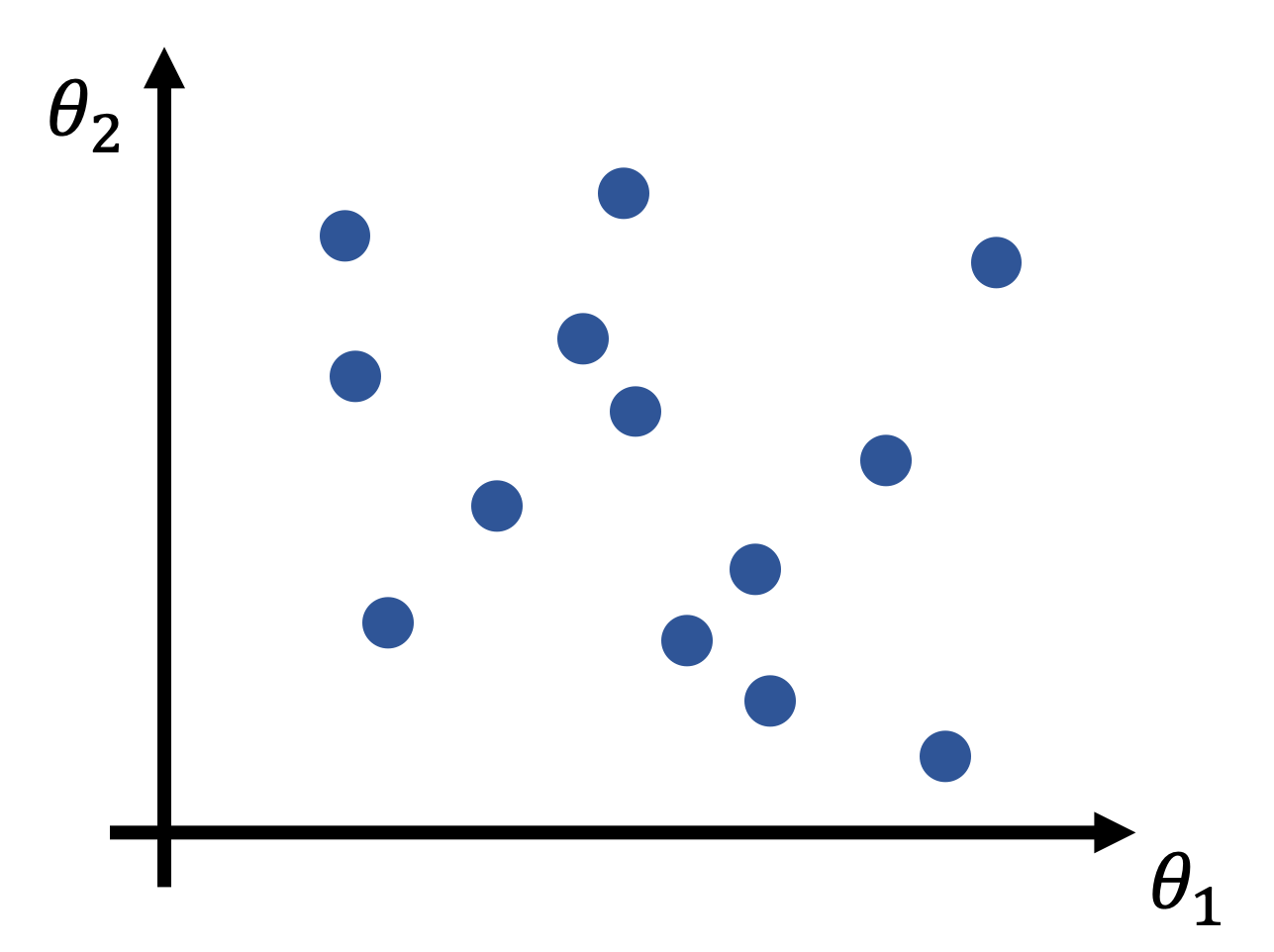}
      }%
      \subfigure[Training of NF models]{%
        \includegraphics[width=0.33\columnwidth,bb=0 0 320 240]{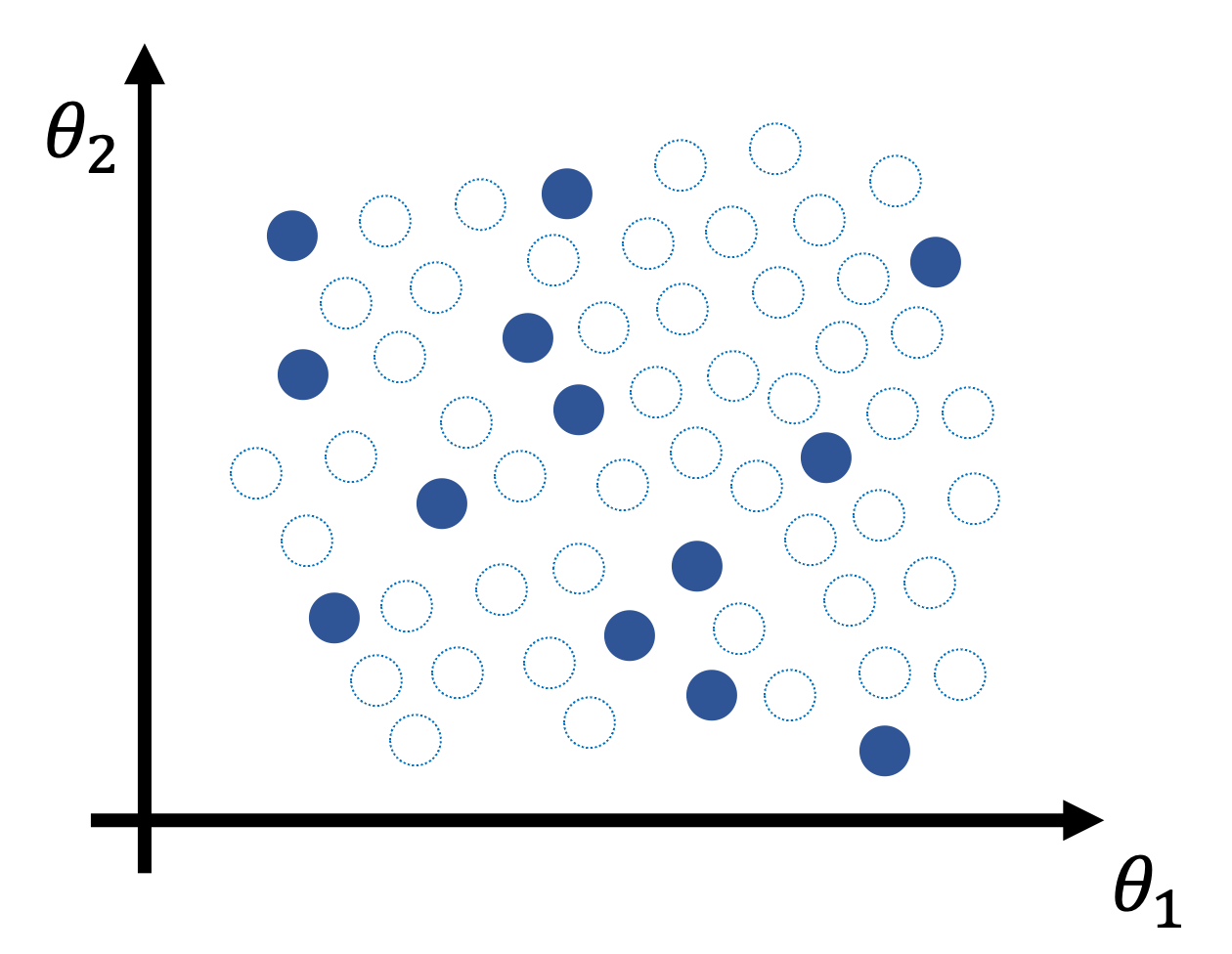}
      }%
      \subfigure[Estimate optimal parameters]{%
        \includegraphics[width=0.33\columnwidth,bb=0 0 320 240]{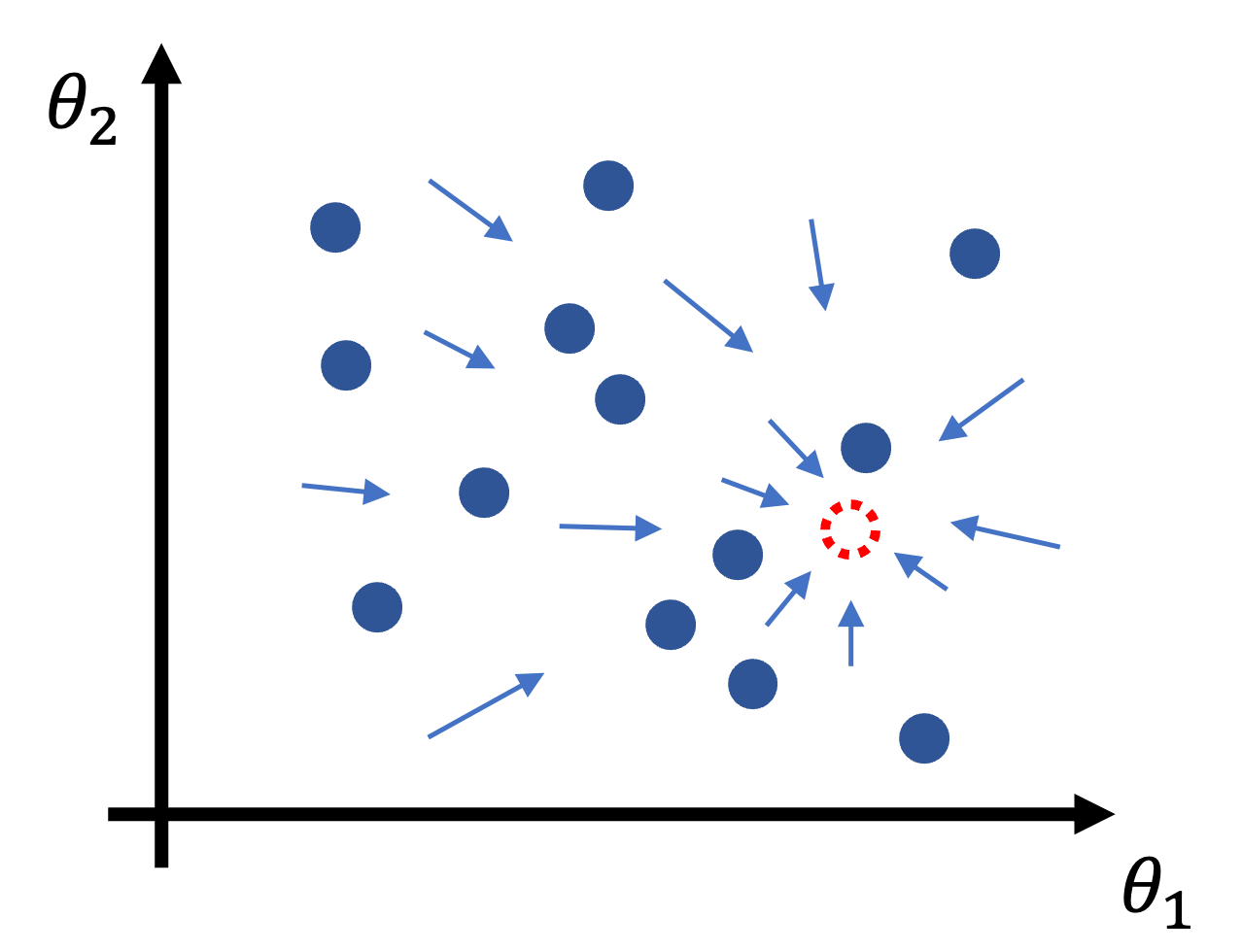}
      }%
    }
    \caption{
      A workflow of the proposed method
    }
    \label{fig:workflow}
  \end{center}
\end{figure}

\begin{description}
\item[Step~1: Generates training samples (Figure~\ref{fig:workflow}(a))] First, the BSM model parameters~($\theta_{\mathrm{BSM}}$) are randomly sampled from the entire phase space, like a random search.
The sampling algorithm is arbitrary, but in this study the parameters for the training samples were sampled from a uniform distribution.
Next, MC samples are generated using the sampled BSM model parameters~($\theta_{\mathrm{BSM}}$).
The sample generation steps depend on the task, but typically include hadronization in Pythia8, detector simulation in Geant4, physics object reconstruction using the experiment's software, and computation of high-level features needed for the physics analysis.
This process can be performed independently for each sampled BSM model parameter, allowing samples to be generated in parallel using distributed computing or other similar resources.

\item[Step~2: Training of normalizing flow models (Figure~\ref{fig:workflow}(b))] Train an NF model~($f_{\mathrm{NF}}(\cdot|\theta_{\mathrm{BSM}})$) using the MC samples generated in Step~1.
Here, a single NF model is trained using all generated samples including the entire phase space, with conditioned BSM signal model parameters~($\theta_{\mathrm{BSM}}$).
After training, the NF model can generate the BSM model distributions with unseen model parameters not used in the training step, i.e., the NF model is able to interpolate in multidimensional space for sample generation.
Since the interpolation performance depends on the complexity of the BSM model distribution and the capability of the NF model, the interpolation capability should be checked after training.

\item[Step~3: Estimate optimal BSM parameter points (Figure~\ref{fig:workflow}(c))] In the physics analysis step, the statistical significance for the existence of the BSM phenomena is evaluated by comparing the experimental data with MC samples, e.g., by performing cut-and-count or fitting the distribution after applying a reasonable selection that rejects the background events while allowing the signal events to pass.
The proposed method can use a metric based on the output of the NF model for the parameter scan other than tuning the selection.
That is, the pdf of the observable variables from the BSM and background events is used to define a likelihood function~($\mathcal{L}(\theta|x_{\mathrm{obs}})$) that is maximized/minimized for the parameter scan.
The advantage of using this metric is that multidimensional distributions of observed data and MC simulated samples can be easily compared without defining a single discriminant feature and tuning the selection depending on signal model parameters.\footnote{Due to the use of multidimensional distributions, systematic uncertainties should be carefully evaluated, as the metric is sensitive to discrepancies between MC and experimental data. This is a future work.}
The gradient of the metric for the BSM model parameters can be calculated using backpropagation. 
Gradient-based optimization methods (e.g., SGD) enables to efficiently find optimal values even in high-dimensional model parameter spaces.
\end{description}

\section{Experiments} \label{sec:experiments}
The proposed method has been applied to two benchmark datasets: the toy dataset~(Sec~\ref{sec:toydata}) and the LHC Olympic 2020 (LHCO2020) benchmark dataset~(Sec~\ref{sec:lhco2020data}).

\subsection{Toy dataset} \label{sec:toydata}
\subsubsection*{Step~1: Generates training samples}
The toy dataset mimics a bump-hunting task, which is a typical problem in collider physics.
The observable~($x$) is one-dimensional and assumes a reconstructed mass~($M_{\mathrm{reco}}$).
It is sampled from analytic functions for signal and background, respectively.
A function for the signal model is the Breit-Wigner function, which has two model parameters, the resonance mass~($M_{\mathrm{pole}}$) and its width~($\Gamma$).
One for the background model is the exponential function, which has a single model parameter~($\tau$).
A training sample was generated with 200k events. The model parameters were uniformly sampled.
The pseudo-sample, which mimics the observed data, was generated with 1k events for the signal and 100k events for the background, with the model parameters~($\theta$) fixed as $(M_{\mathrm{pole}}, \Gamma, \tau) = (91.2, 2.5, 10^{-2})$.

\subsubsection*{Step~2: Training of normalizing flow models}
\begin{figure}[htbp]
  \begin{center}
    {
      \subfigure[Signal]{%
        \includegraphics[width=0.48\linewidth,bb=0 0 528 396]{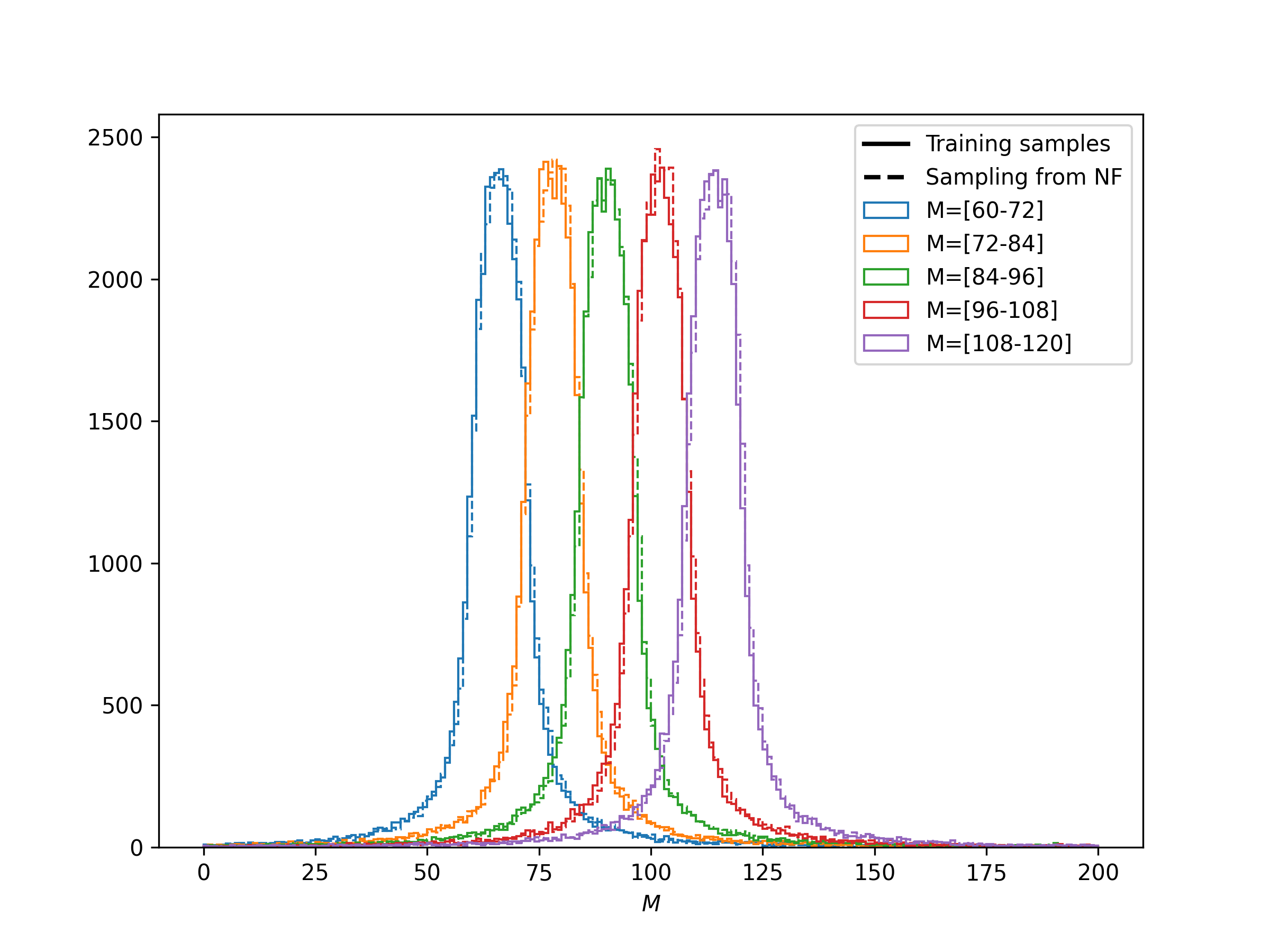}
      }%
      \subfigure[Background]{%
        \includegraphics[width=0.48\linewidth,bb=0 0 528 396]{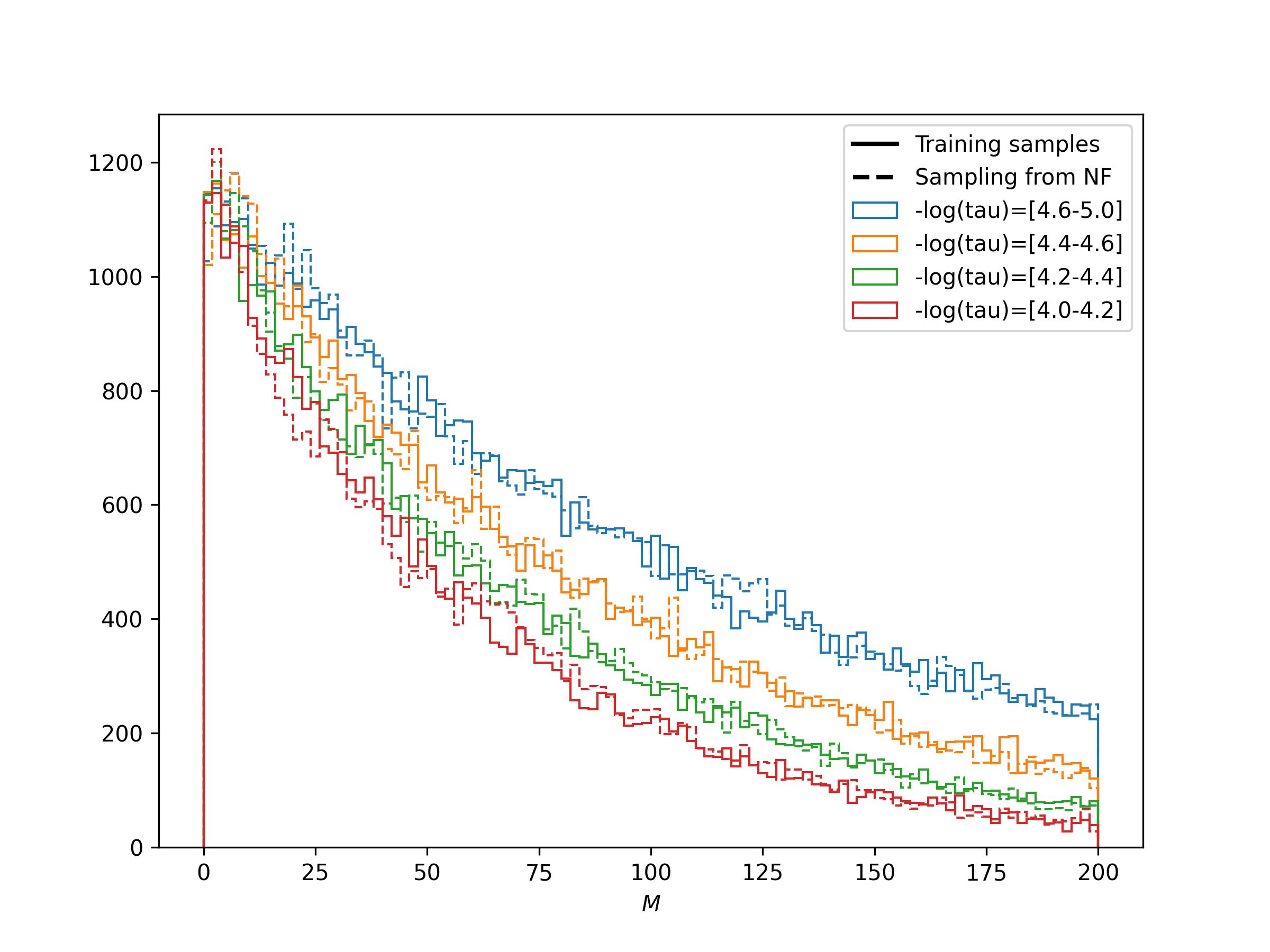}
      }%
    }%
    \caption{
      Fitting results for (a) signal distributions and (b) background distributions.
      The distributions of the training samples are shown as a solid line, and the distributions sampled from the trained NF model are shown as a dashed line, colored by the model parameters~($M$ for signal, and $-\log(\tau)$ for background).
      As the model parameters are sampled uniformly, the distributions are colored by each interval for the model parameters~(e.g. $M$=[60-72] means $60 < M < 72$).
      A single signal/background NF model is used to model the distributions generated by all signal/background models with different model parameters.
    }
    \label{fig:toy_fitting}
  \end{center}
\end{figure}

SplineFlows~\cite{durkan2019neural} was used as the NF model.
SplineFlows uses a spline function in each transformation step, resulting in strong representation capability.
The loss function for training the NF model is defined as a negative log probability~($-\log p_{\mathrm{NF}}(x|\theta)$).
Figure~\ref{fig:toy_fitting} shows the distributions of the observables of the training samples and one sampled from the trained NF model, colored by different model parameters.
Here, only one NF model is used for signal and background respectively, instead of defining several models for each model parameter.
The NF model has the ability to model the distribution for multiple model parameters.

\subsubsection*{Step~3: Estimate optimal BSM parameter points}
\begin{figure}[htbp]
  \begin{center}
    {
      \subfigure[NLL ($\Delta (-\log \mathcal{L})$)]{%
        \includegraphics[width=0.48\linewidth,bb=0 0 528 396]{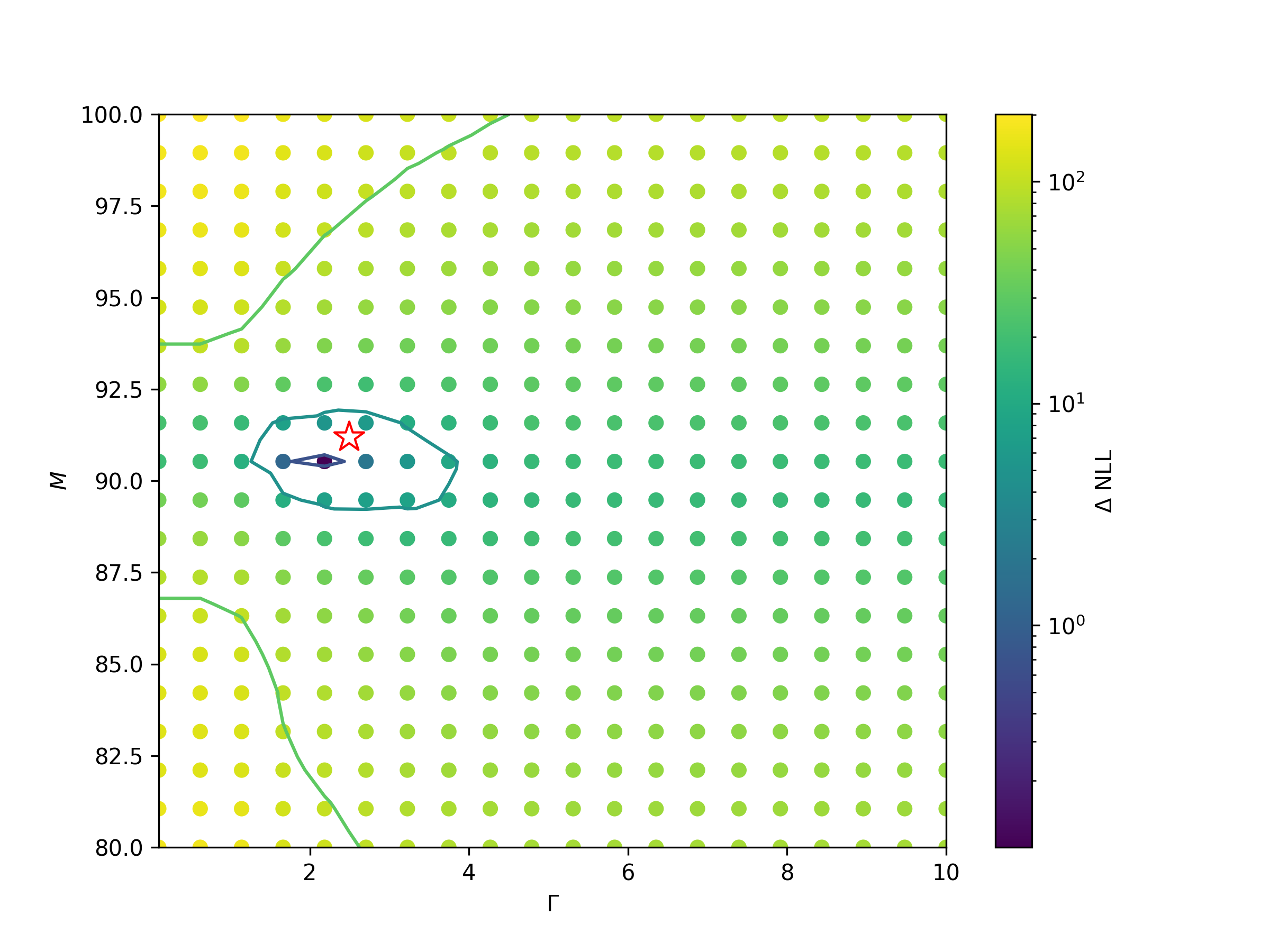}
      }%
      \subfigure[Gradients ($\nabla_{\theta}(-\log \mathcal{L})$)]{%
        \includegraphics[width=0.48\linewidth,bb=0 0 528 396]{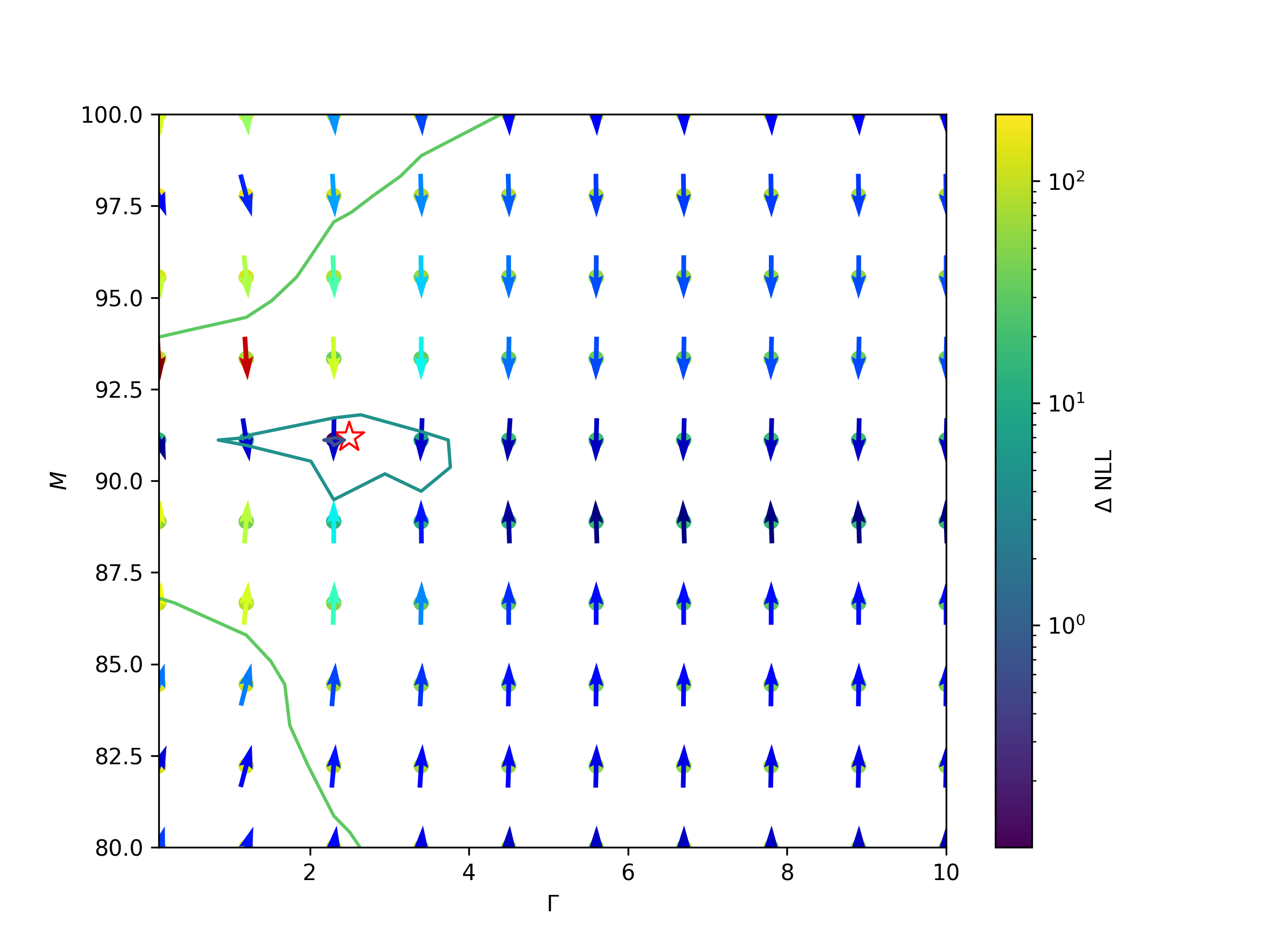}
      }%
    }
    \caption{
      (a) The NLL values and (b) the gradients in the signal model parameter space~($(M_{\mathrm{pole}}, \Gamma)$).
      The red star is the truth model parameter points.
      The arrows in (b) represent gradients.
    }
    \label{fig:toydata_nll}
  \end{center}
\end{figure}

The likelihood function used for the parameter scan is defined as 
\begin{equation*}
  \mathcal{L}(\{ M_{\mathrm{pole}}, \Gamma, \tau \} | M_{\mathrm{reco}}) 
  = \frac{n_{\mathrm{sig}}}{n_{\mathrm{sig}} + n_{\mathrm{bg}}} 
  \cdot 
  p_{\mathrm{NF}}^{\mathrm{sig}}(M_{\mathrm{reco}}|\{ M_{\mathrm{pole}}, \Gamma \}) 
  + \frac{n_{\mathrm{bg}}}{n_{\mathrm{sig}} + n_{\mathrm{bg}}} 
  \cdot 
  p_{\mathrm{NF}}^{\mathrm{bg}}(M_{\mathrm{reco}}|\tau),
\end{equation*}
where $n_{\mathrm{sig}}$ and $n_{\mathrm{bg}}$ are the number of signal events and the number of background events, respectively, and $p_{\mathrm{NF}}^{\mathrm{sig/bg}}(M_{\mathrm{reco}}|\theta)$ is a probability inferred by the NF model.
$n_{\mathrm{sig}}$ and $n_{\mathrm{bg}}$ can be treated as floating parameters, but they are fixed here for simplicity.
The negative log likelihood~(NLL) is computed for all the pseudodata~($-\log \mathcal{L}(\theta | M_{i, \mathrm{reco}})$), and summed~($\sum_{M_i \in \mathrm{pseudo\ data}} -\log \mathcal{L}(\theta | M_{i, \mathrm{reco}})$) to construct an objective function to be optimized.
Figure~\ref{fig:toydata_nll}~(a) shows the NLL values in the signal model parameter space~($(M_{\mathrm{pole}}, \Gamma)$).
The $z$-axis is the $\Delta$~NLL, the difference from the smallest NLL obtained\footnote{A tiny non-zero value is added for logarithmic visualization, i.e. $\log (\mathrm{NLL} - \mathrm{NLL}_{\mathrm{min}} + \epsilon)$ is plotted }.
The red star indicates the target model parameter points used to generate the pseudo data. It is close to the minimum NLL point within the grid size.
Figure~\ref{fig:toydata_nll}~(b) shows the gradient of the NLL in the model parameter spaces~($\nabla_{\theta}(-\log \mathcal{L})$). The direction of the arrow indicates the direction of the gradient, and the color of the arrow indicates the magnitude of the gradient.
In this task, the gradient along the $\Gamma$ axis is small, and the gradient direction aligns with the $M_{\mathrm{pole}}$ direction in the large $\Gamma$ region.
This non-uniformity can be mitigated by using some optimization methods with momentum technique.
In the small $\Gamma$ regions, the gradient points to the true minimum, and it is expected that the gradient-based optimization method can be used for the parameter scan.

\subsection{LHC Olympic 2020 dataset} \label{sec:lhco2020data}
The other application results are shown using a more complex dataset, the LHC Olympic 2020 (LHCO 2020) benchmark dataset~\cite{Kasieczka_2021}.

\subsubsection*{Step~1: Generates training samples}
LHCO 2020 is a benchmark dataset prepared for a machine learning anomaly detection competition held in 2020.
This dataset contains the signal process where the $Z^{'}$ boson decays into two unknown heavy bosons, $X$ and $Y$ bosons, and each of them decays into a quark pair.
Since the $X$/$Y$ boson has a large momentum, the quark pair is boosted, resulting in the merging of two remnants of each quark into a single observed jet, the so-called large-R jet. As a result, two large-R jets are observed as a signal event.
This dataset contains a set of 4-vectors of all stable particles in the collision events, but 
high-level features were used for simplicity by reconstructing the jets using the FastJet library.
The definitions of the high-level features used as input variables and preselections are based on the ANODE~\cite{PhysRevD.101.075042}.
The five high-level features are the leading jet mass~($M_{J_1}$), the difference between leading and subleading jet mass~($M_{J_1} - M_{J_2}$), the jet substructure variables for leading and subleading jets~($\tau_{J_{1},21}$ and $\tau_{J_{2},21}$), and the dijet mass~($M_{J_{1}J_{2}}$), where $M_{J_1}$, $M_{J_2}$, and $M_{J_{1}J_{2}}$ are highly correlated with the $X$ boson mass~($M_{X})$, the $Y$ boson mass~($M_{Y})$, and the $Z^{'}$ boson mass~($M_{Z^{'}})$, respectively.
Since the signal model parameters are the masses of three boson~($M_{X}$, $M_{Y}$, $M_{Z^{'}}$), the dimension of this task is 3 for the signal phase space and 5 for the observable variables.

The benchmark dataset was provided with $Z^{'}$, $X$, and $Y$ masses of 3500, 500, and 100 GeV, respectively.
For this study, the dataset was extended with additional model parameters.
The signal parameters were chosen according to the grid: $m_{Z^{'}}$ = [3000, 3500, 4000, 4500] GeV, $m_{X}$ = [250, 500, 750, 1000] GeV, and $m_{Y}$ = [50, 100, 150, 200, 250, 300] GeV, where only combinations satisfying the conditions of $m_{X} - m_{Y} > 100~\mathrm{GeV}$ were generated as training samples.
The samples at each grid point were processed using the same event generator configuration as the original benchmark dataset. 
The total number of generated signal events was 583k.
The background samples were the original one, and the number of events is 91k.

All 3 model parameters and 5 observables were normalized by linear transformation and scaled to the range of 0 to 1.

\subsubsection*{Step~2: Training of normalizing flow models}
Masked Autoregressive Flow~(MAF)~\cite{papamakarios2018masked} was used as the NF model.
As in Section~\ref{sec:toydata}, a single NF model was defined and trained for each signal and background distribution with all model parameters.

\begin{figure}[htbp]
  \begin{center}
    {
      \subfigure[$M_{J_{1}J_{2}}$]{%
        \includegraphics[width=0.48\linewidth,bb=0 0 528 396]{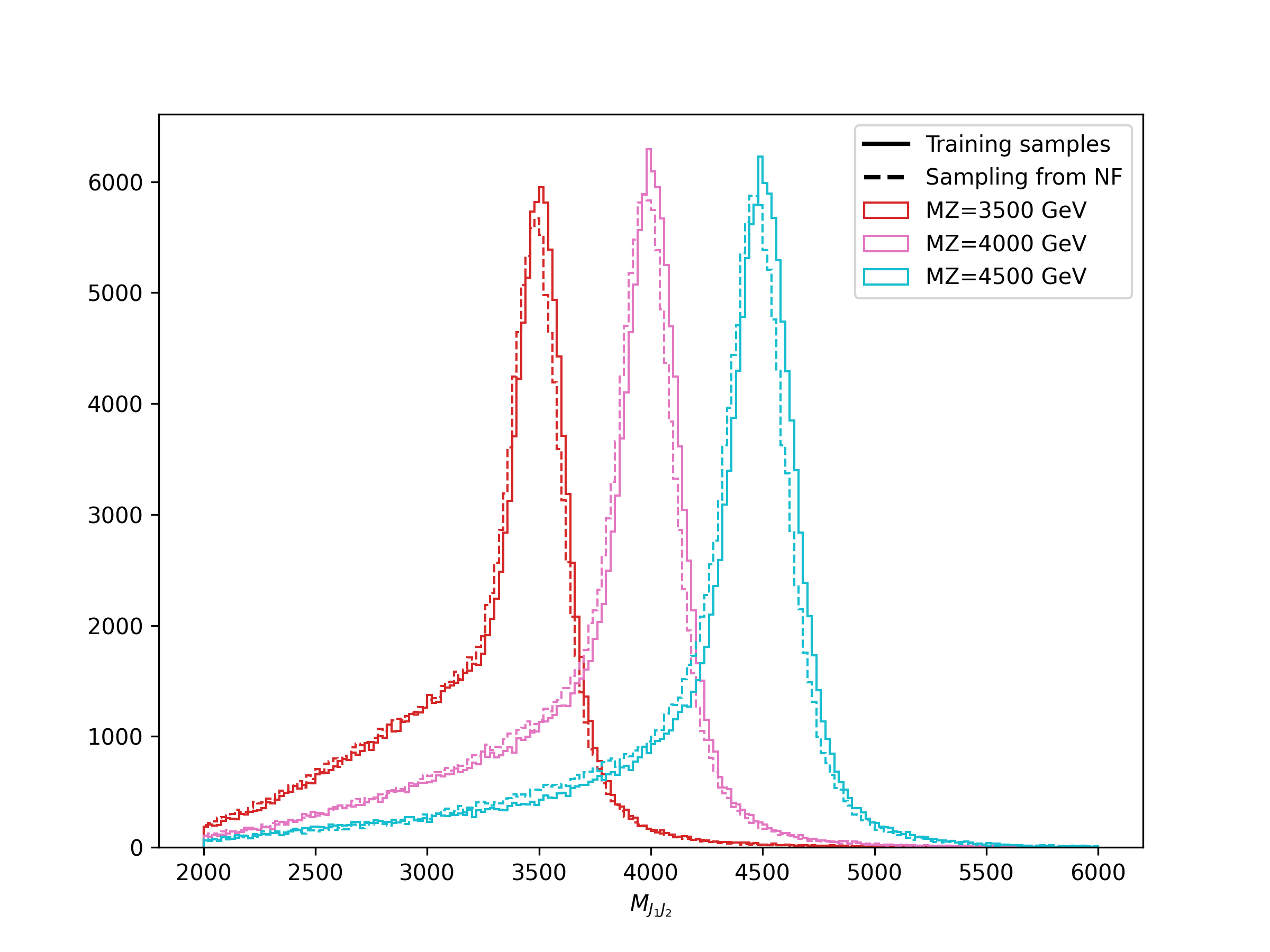}
      }%
      \subfigure[$M_{J_{1}}$]{%
        \includegraphics[width=0.48\linewidth,bb=0 0 528 396]{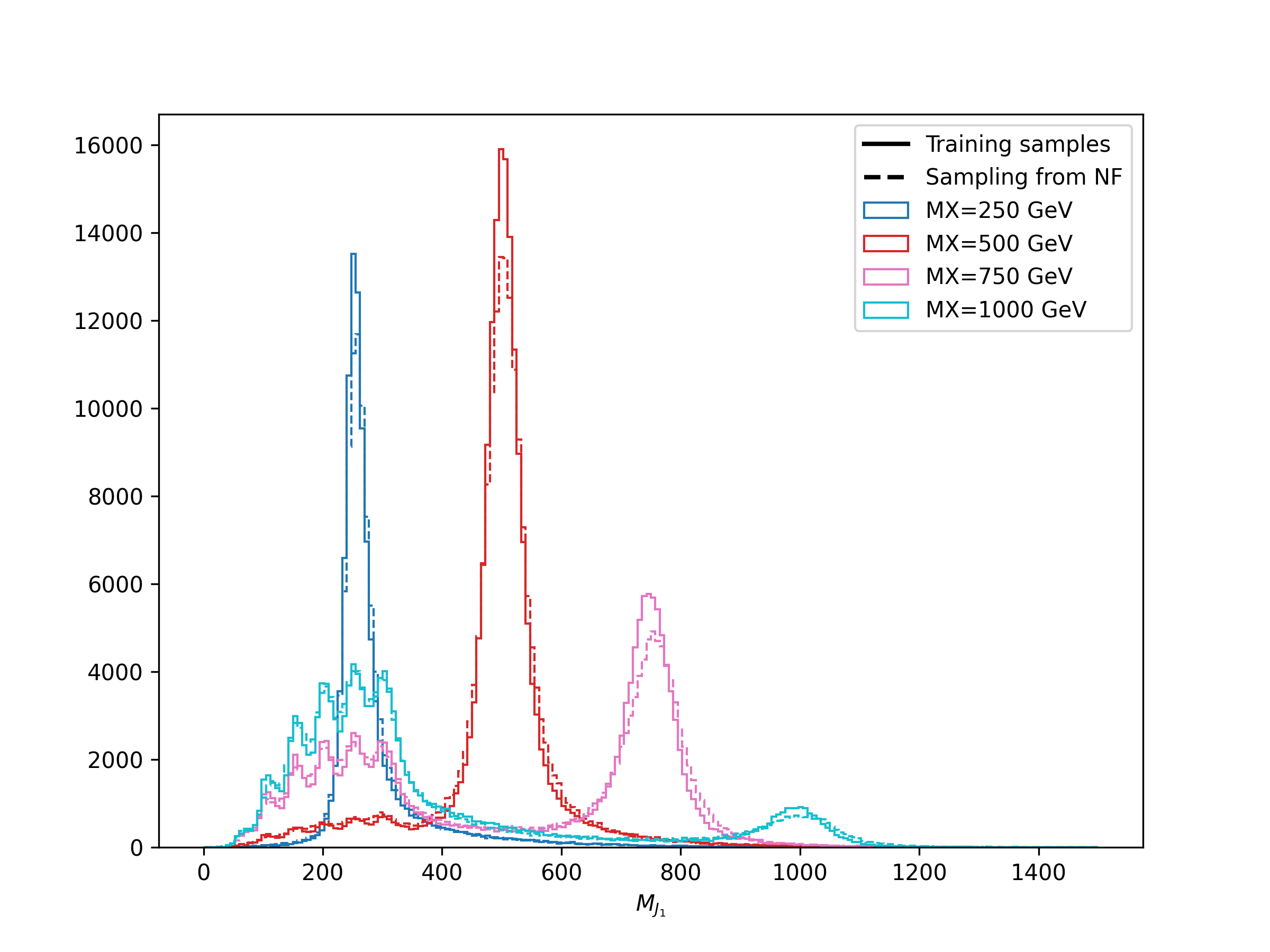}
      }%
    }
    \caption{
      Fitting results for (a) dijet mass~($M_{J_{1}J_{2}}$) and (b) the leading jet mass~($M_{J_{1}}$) distributions.
      The distributions of the training samples are shown as a solid line, and the distributions sampled from the trained NF model are shown as a dashed line, colored by the model parameters.
      A single signal NF model is used to model the distributions generated by all signal models with different model parameters.
    }
    \label{fig:lhco2020_fitting}
  \end{center}
\end{figure}

Figure~\ref{fig:lhco2020_fitting} shows the observable distribution of the training sample and the distribution sampled by the trained NF model, colored by the different model parameters.
Each distribution is well modeled by the NF model for all signal model parameters.
A strange structure can be seen in Figure~\ref{fig:lhco2020_fitting}~(b).
It comes from the swapping of the leading and subleading jets, i.e. the $Y$ boson mass peak appears in the $M_{J_{1}}$ distribution. For such a complex distribution, the NF model is able to model the distribution well.

\begin{figure}[htbp]
  \begin{center}
    {
      \subfigure[Interpolation]{%
        \includegraphics[width=0.48\linewidth,bb=0 0 528 396]{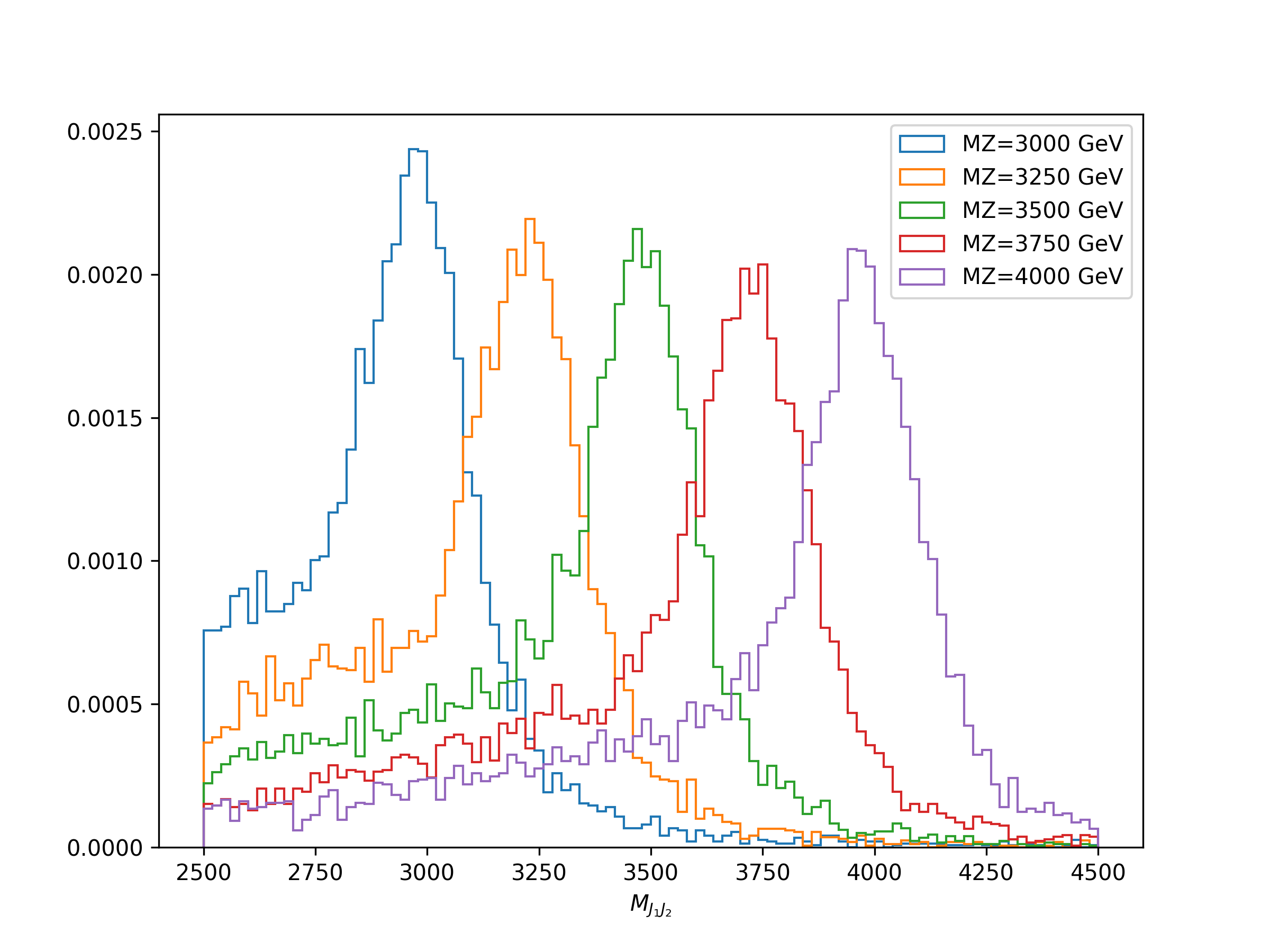}
      }%
      \subfigure[Extrapolation]{%
        \includegraphics[width=0.48\linewidth,bb=0 0 528 396]{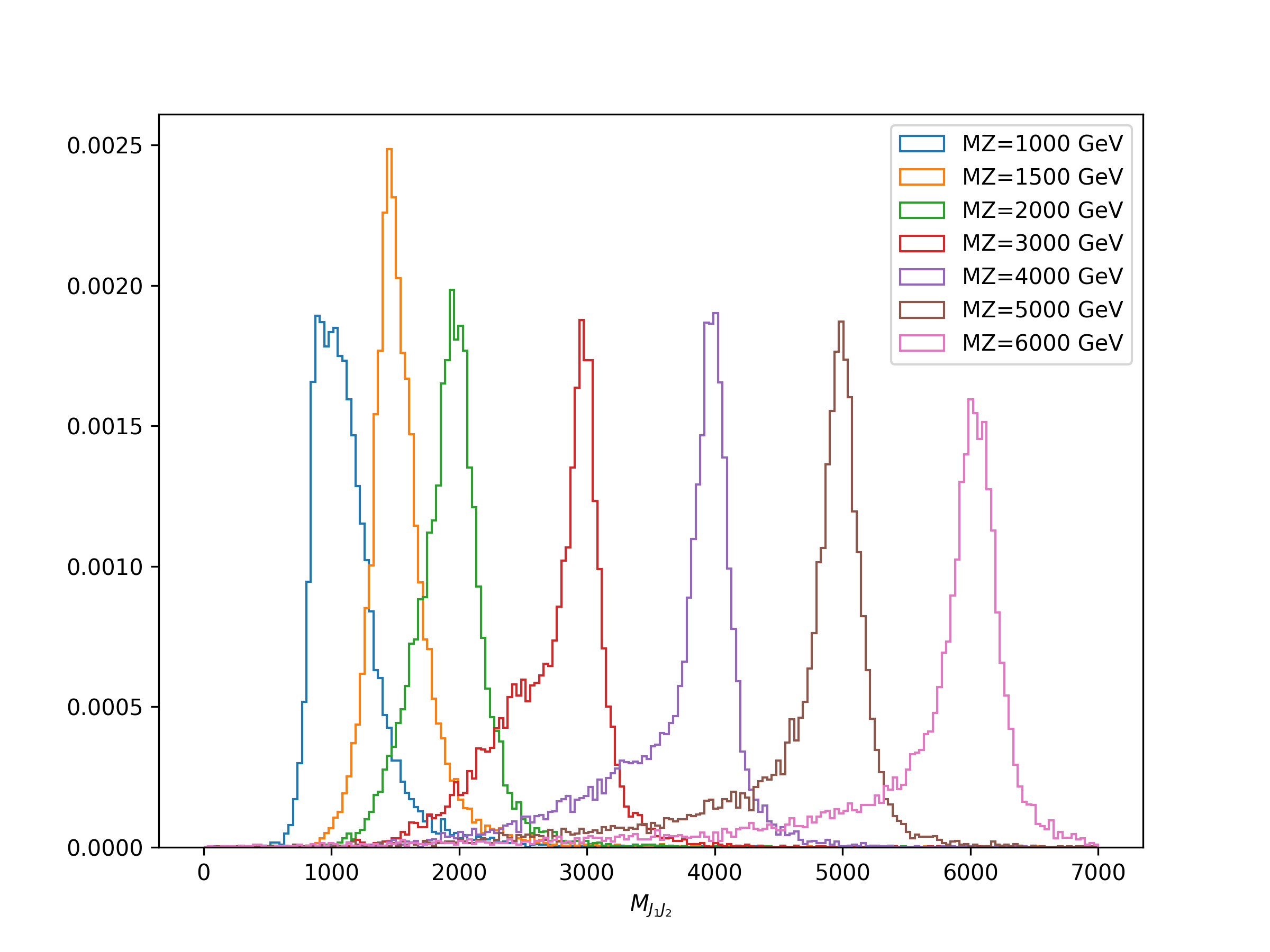}
      }%
    }
    \caption{
      Dijet mass~($M_{J_{1}J_{2}}$) distribution sampled by the trained NF models.
      Model parameters used in the training samples were $m_{Z^{'}}$ = [3000, 3500, 4000, 4500] GeV.
      All lines in (a) and (b), including the model parameter above, are sampled by the NF model.
    }
    \label{fig:lhco2020_interpolation}
  \end{center}
\end{figure}

Figure~\ref{fig:lhco2020_interpolation} shows the $M_{J_{1}J_{2}}$ distribution sampled by the NF model, colored by the signal model parameters, some of which are not used in the training sample.
The training sample was generated as $m_{Z^{'}}$ = [3000, 3500, 4000, 4500] GeV, and the distributions of the other model parameters were generated by the NF model by interpolation.
In Figure~\ref{fig:lhco2020_interpolation}~(a), the orange and red lines are the interpolated results, indicating that the NF model can interpolate correctly.
In Figure~\ref{fig:lhco2020_interpolation}~(b), all lines except the red and purple lines are the extrapolated results, also indicating that the NF model can sample a reasonable distribution.

\subsubsection*{Step~3: Estimate optimal BSM parameter points}
The likelihood function is defined as in the case of toy data.
\begin{eqnarray*}
  \mathcal{L}(\theta_{\mathrm{sig}}| x) 
  &=& \frac{n_{\mathrm{sig}}}{n_{\mathrm{sig}} + n_{\mathrm{bg}}} 
  \cdot 
  p_{\mathrm{NF}}^{\mathrm{sig}}(x|\theta_{\mathrm{sig}}) 
  + \frac{n_{\mathrm{bg}}}{n_{\mathrm{sig}} + n_{\mathrm{bg}}} 
  \cdot 
  p_{\mathrm{NF}}^{\mathrm{bg}}(x) \\
  x &=& \{M_{J_1}, M_{J_1} - M_{J_2}, \tau_{J_{1},21}, \tau_{J_{2},21}, M_{J_{1}J_{2}} \} \\
  \theta_{\mathrm{sig}} &=& \{M_{Z^{'}}, M_{X}, M_{Y} \} ,
\end{eqnarray*}
where the no conditional parameters~($\theta_{\mathrm{bg}}$) are defined for the NF model for background samples. The model parameters for background samples, such as hadronization parameters, generator difference, etc., can be treated as the background model parameters. This is a future work.

\begin{figure}[htbp]
  \begin{center}
    {
      \subfigure[$M_{J_{1}J_{2}}$]{%
        \includegraphics[width=0.33\linewidth,bb=0 0 528 396]{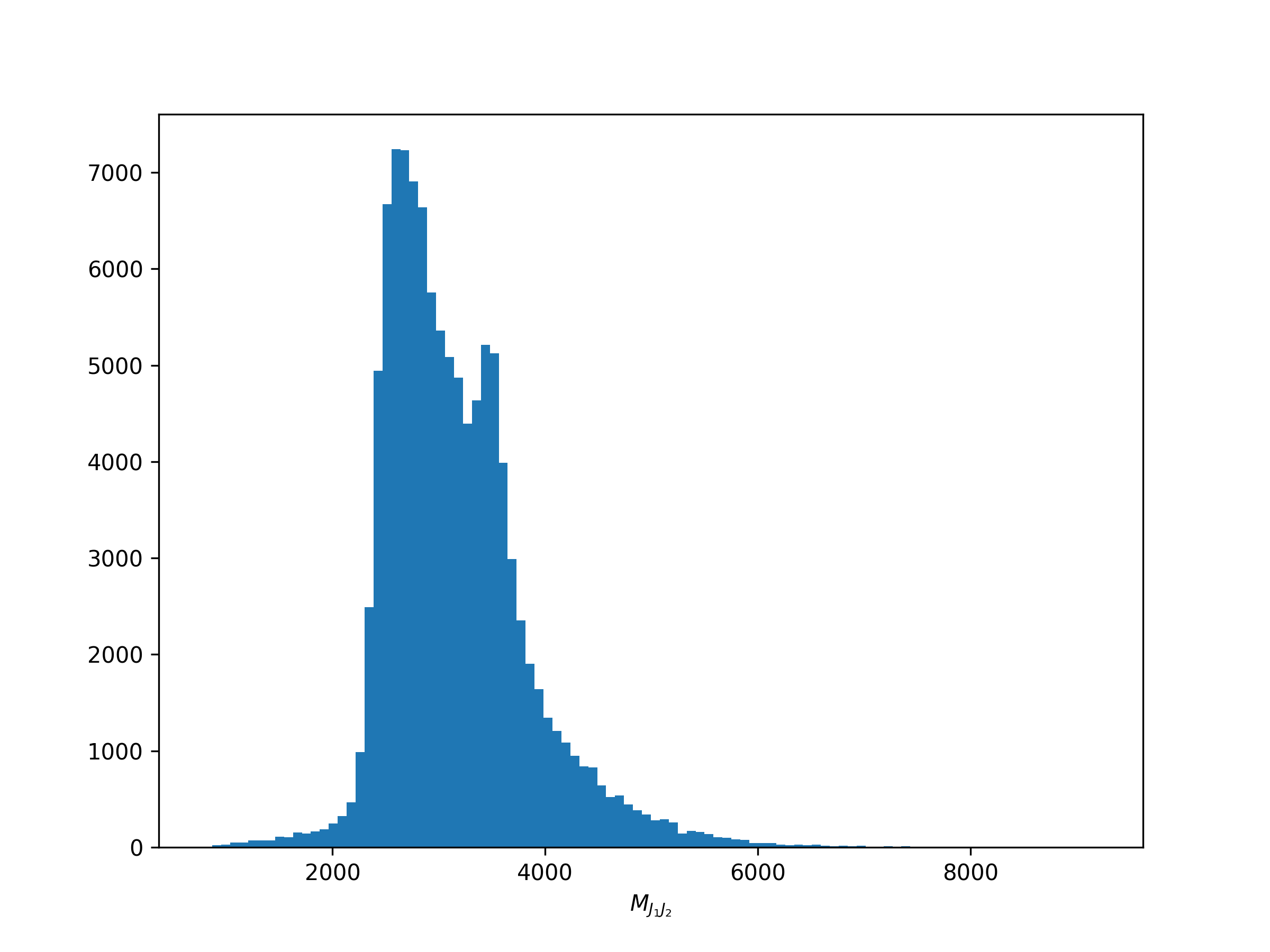}
      }%
      \subfigure[$M_{J_1}$]{%
        \includegraphics[width=0.33\linewidth,bb=0 0 528 396]{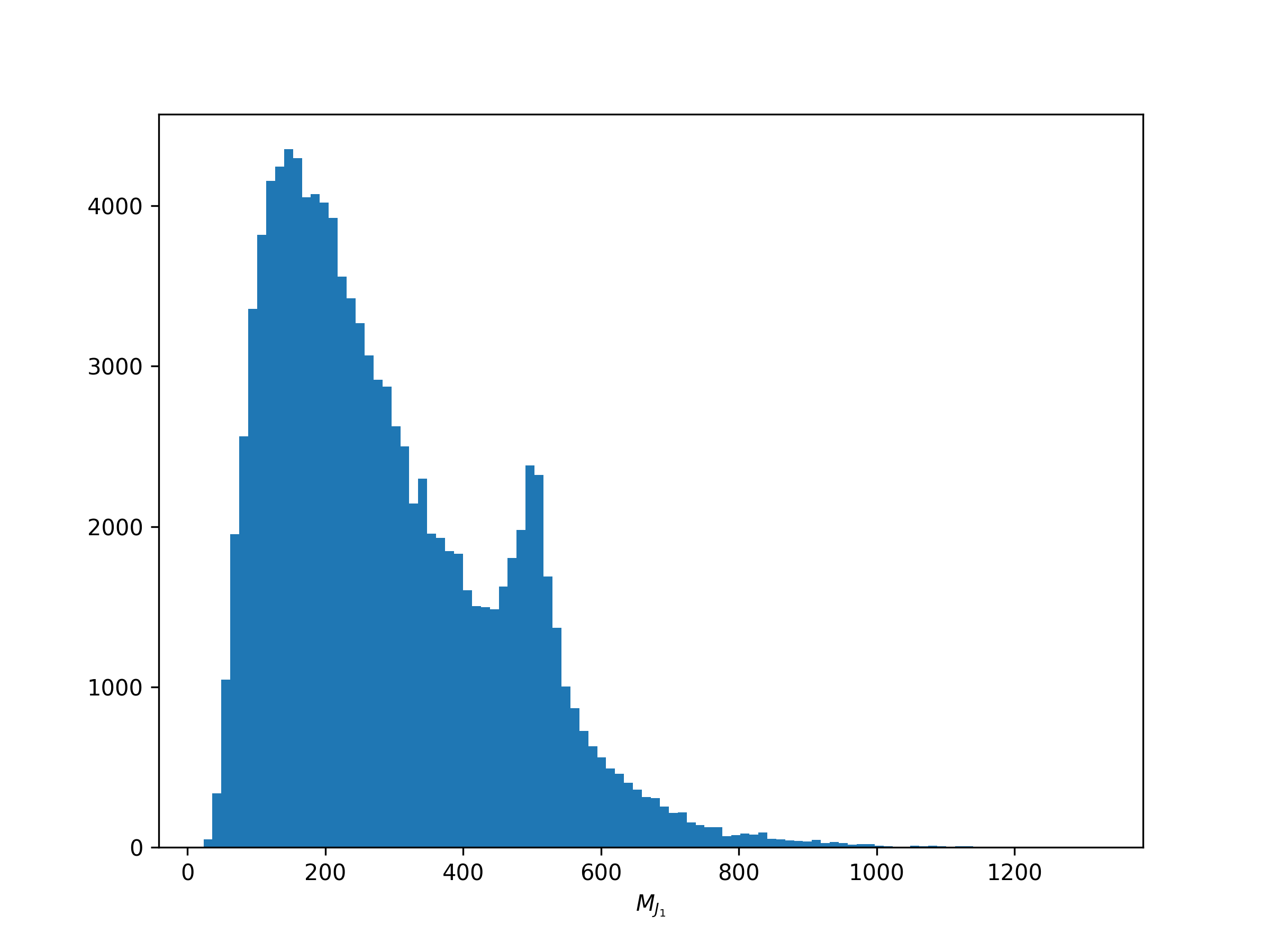}
      }%
      \subfigure[$M_{J_1} - M_{J_2}$]{%
        \includegraphics[width=0.33\linewidth,bb=0 0 528 396]{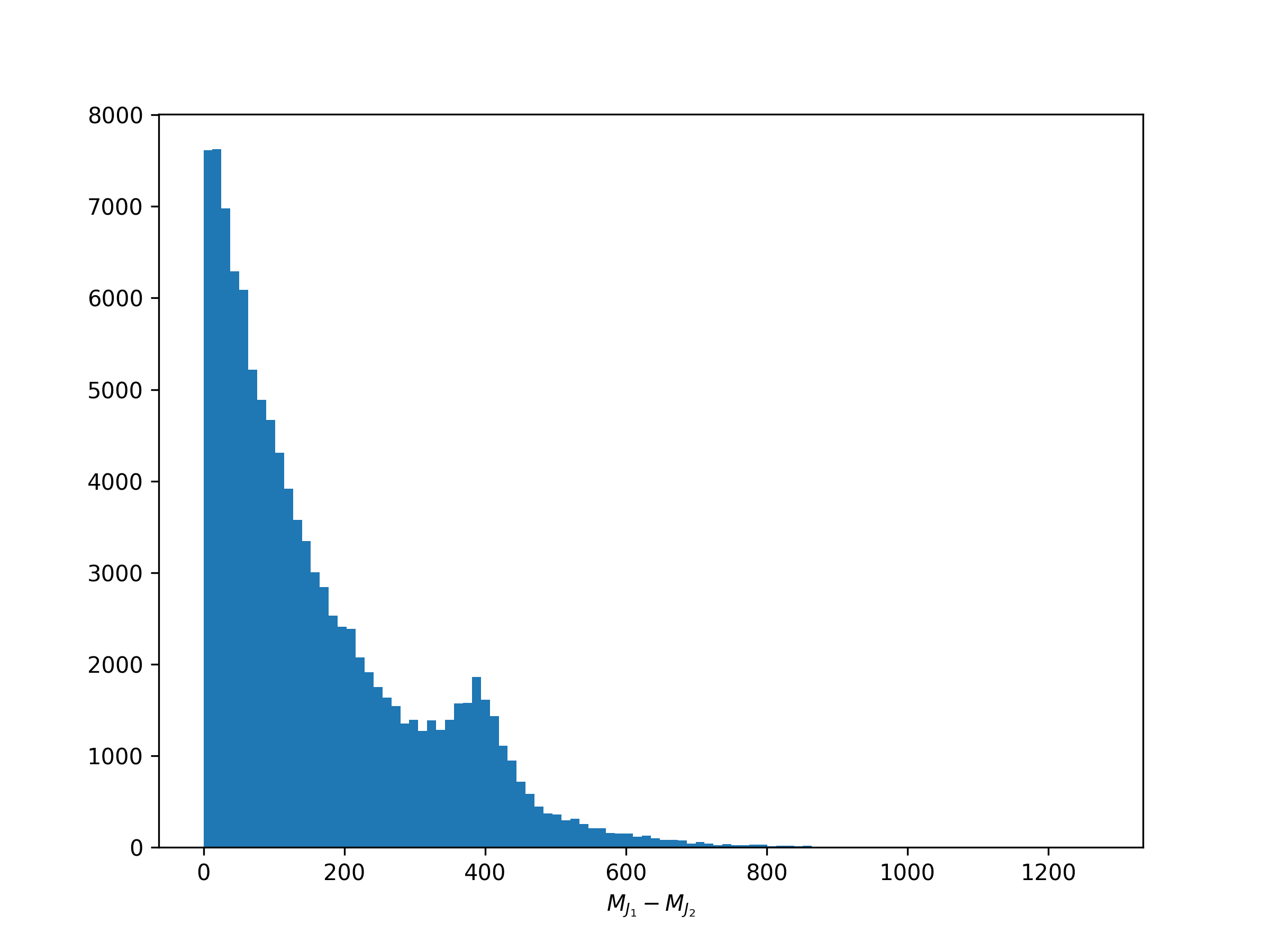}
      }%
    }
    \caption{
      The observable distribution in the pseudo dataset.
      The signal events are located at ($M_{J_{1}J_{2}}$, $M_{J_1}$, $M_{J_1} - M_{J_2}$) = (3500, 500, 400) GeV.
    }
    \label{fig:lhco2020_pseudodataset}
  \end{center}
\end{figure}

The pseudo dataset mimicking the observed data was bootstrapped from the samples with ($M_{Z^{'}}$, $M_{X}$, $M_{Y}$) = (3500, 500, 100) GeV, with the number of events was set to 10k for the signal and 100k for the background.
The observable distributions of the pseudo dataset are shown in Figure~\ref{fig:lhco2020_pseudodataset}.

\begin{figure}[htbp]
  \begin{center}
    {
      \subfigure[NLL ($\Delta (-\log \mathcal{L})$)]{%
        \includegraphics[width=0.48\linewidth,bb=0 0 528 396]{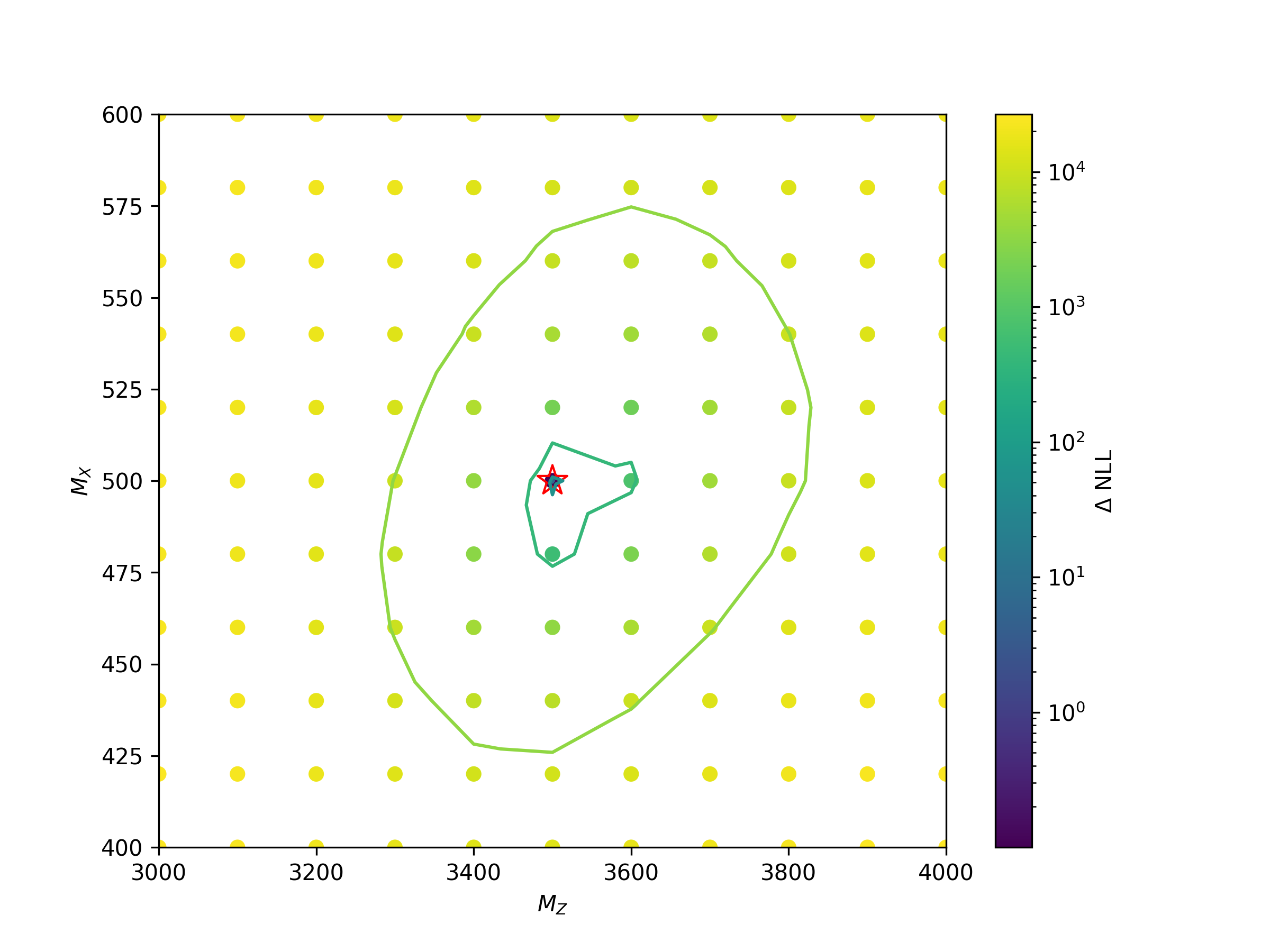}
      }%
      \subfigure[Gradients ($\nabla_{\theta}(-\log \mathcal{L})$)]{%
        \includegraphics[width=0.48\linewidth,bb=0 0 528 396]{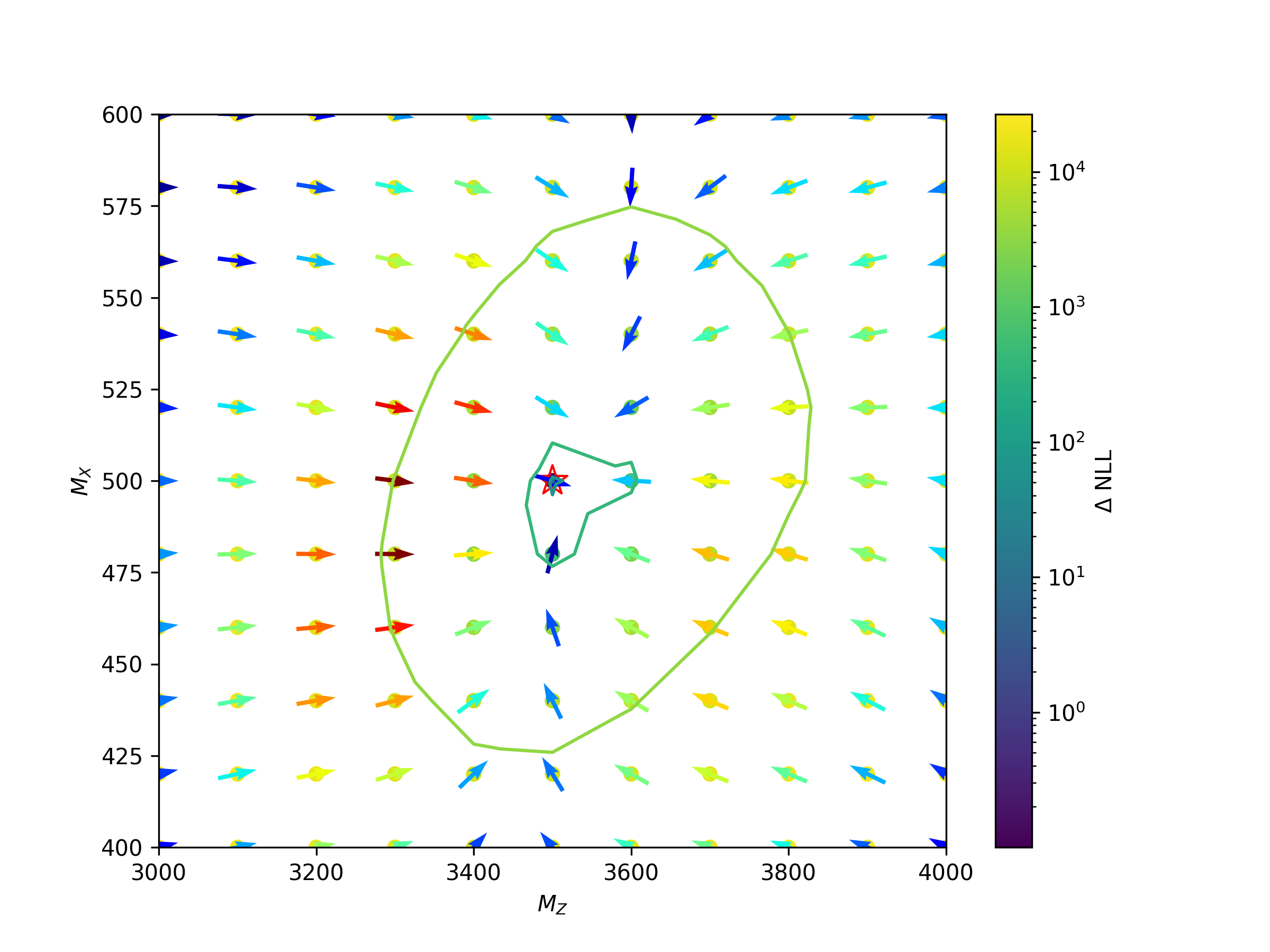}
      }%
    }
    \caption{
      (a) The NLL values and (b) the gradients in the signal model parameter space~($(M_{Z^{'}}, M_{X})$).
      The red star is the truth model parameter points.
      The arrows in (b) represent gradients.
    }
    \label{fig:lhco2020_nll}
  \end{center}
\end{figure}

Figure~\ref{fig:lhco2020_nll} shows the $\Delta$ NLL distribution and the distribution of the gradient, using the same plot style as Figure~\ref{fig:toydata_nll}.
Even with high-dimensional observables and complex feature distributions, the gradient for the target function is correctly evaluated.

\section{Conclusion}
This paper proposes an efficient signal model parameter scan technique based on normalizing flow. 
This technique is demonstrated on toy dataset and LHC Olympic 2020 benchmark dataset, and confirmed that the normalizing flow model has good capabilities for modeling complex distributions and interpolating signal model parameter spaces, and the gradient of NLL can be evaluated quickly by backpropagation.
It is expected that this technique can be extended to higher-dimensional data; higher-dimensional observables~($x$), e.g. the set of particle four-vectors, and higher-dimensional model parameters~($\theta$), e.g. phenomenological MSSM.

\section*{Acknowledgement}
This work was supported by JSPS KAKENHI Grant Numbers JP21K13936 and JP22H05113, and partially supported by Institute of AI and Beyond for the University of Tokyo.

\bibliography{main}

\end{document}